\documentclass{svproc}
\usepackage{url}

\usepackage{caption}
\usepackage{multirow}
\usepackage{bbding}
\usepackage{todonotes}
\usepackage{hyperref}
\usepackage{ulem}

\begin{document}

\mainmatter              % start of a contribution
\title{Comparison of System Call Representations for Intrusion Detection}
\titlerunning{System Call Representation}  % abbreviated title (for running head)
\author{Sarah Wunderlich\inst{1}\Envelope \and Markus Ring\inst{1}
Dieter Landes\inst{1} \and Andreas Hotho\inst{2}}
\authorrunning{Sarah Wunderlich et al.} % abbreviated author list (for running head)
\tocauthor{Sarah Wunderlich, Markus Ring, Dieter Landes and Andreas Hotho}
\institute{Coburg University of Applied Sciences and Arts, \\Coburg, Germany\\
\email{sarah.wunderlich@hs-coburg.de},\\ WWW home page:
\texttt{http://www.hs-coburg.de/cidds}
\and
Data Mining and Information Retrieval Group,\\
University of W\"urzburg,
\\W\"urzburg, Germany}

\maketitle              % typeset the title of the contribution

\begin{abstract}
Over the years, artificial neural networks have been applied successfully in many areas including IT security. 
Yet, neural networks can only process continuous input data. 
This is particularly challenging for security-related non-continuous data like system calls.
This work focuses on four different options to preprocess sequences of system calls so that they can be processed by neural networks.
These input options are based on one-hot encoding and learning word2vec or GloVe representations of system calls.
As an additional option, we analyze if the mapping of system calls to their respective kernel modules is an adequate generalization step for (a) replacing system calls or (b) enhancing system call data with additional information regarding their context.
However, when performing such preprocessing steps it is important to ensure that no relevant information is lost during the process.
The overall objective of system call based intrusion detection is to categorize sequences of system calls as benign or malicious behavior.
Therefore, this scenario is used to evaluate the different input options as a classification task.
The results show, that each of the four different methods is a valid option when preprocessing input data, but the use of kernel modules only is not recommended because too much information is being lost during the mapping process.
\keywords{HIDS, System Calls, Intrusion Detection, LSTMs}
\end{abstract}

\section{Introduction}
In the age of digitization, many process flows involving personal or otherwise critical data have been digitized.
Hence, it is more important than ever to protect data against unauthorized access and a lot of research on intrusion detection has been done over the past few years.
An obvious approach for this task is to monitor the system calls of different processes running on a host.
Since standard application programs are running in user mode, they are not allowed to access the resources of an operating system on their own.
Even for simple activities (like reading or writing a file), programs need to make a request to the operating system's kernel in the form of system calls.
Hence, if a program is exploited, every action the exploit takes will also be mirrored within the system call trace of the program.
Consequently, system calls have widely been used as data source for security-critical events in intrusion detection \cite{creech2014semantic,forrest1996sense,hofmeyr1998intrusion,kim2016lstm,kolosnjaji2016deep,sharma2007intrusion}.
A simplified approach is using the kernel modules to which the system calls belong rather than the system calls \cite{murtaza2015trace}.

\textbf{Problem Setting.} 
The increasing interest in (deep) neural networks has also reached the area of IT security in recent years.
Neural networks, however, can only process continuous input data. 
Yet, many security-related data like system calls are non-continuous data which constitutes a major limitation to the application of neural networks in this area. 
In particular, neural networks cannot be directly applied to system calls as these do not conform to the expected input formats.

\textbf{Objective.} 
We intend to (pre--)process system calls in such a way that they can be analyzed through neural networks.
Preprocessing can be accomplished in many ways. 
Complex structures like language demand a more sophisticated preprocessing step since, e.g., context information can play a decisive role for understanding a sentence.
The field of natural language processing (NLP) has different approaches for inserting context into a word. 
Since logs of system calls are basically textual data where temporal ordering is relevant, we want to investigate the suitability of different approaches from NLP to transform system calls into continuous input data for neural networks.

\textbf{Approach and Contribution.}
Benign and malicious behavior of processes can be analyzed by looking at sequences of system calls. 
Following this line, this work explores Long Short Term Memory Networks (LSTMs) for processing sequences of system calls. 
LSTMs are a type of neural networks which are able to process sequence data \cite{hochreiter1997long}. 
As mentioned above, preprocessing system calls such that they can be processed by neural networks is a crucial issue.
In particular, it should be ensured that no relevant information is lost during the preprocessing step. 
We examine four preprocessing approaches from NLP, namely (1) one-hot encoding, (2) expanding the original network structure by an additional embedding layer and learning (3) word2vec and (4) GloVe representations of system calls prior to the classification task.
Evaluation uses the ADFA-LD dataset \cite{creech2013generation}.

%Aside from the original system calls, we analyze if a generalization step is advantageous by mapping the system calls to their respective kernel modules they are defined in.
As an alternative to using the original system calls, several generalizations might be applicable.
For instance, system calls can be mapped to the kernel modules in which they are defined in.
The four different representations used on the raw system call data are also applied to this kernel module representation. 
Because the respective kernel modules may contain additional information about the relationship between different system calls, a combination of system calls and kernel modules is also examined as input.

The paper's main contribution is a systematic evaluation of different preprocessing methods for system calls such that they can be used in LSTMs for detecting security critical events (malicious behavior). 
To the best of our knowledge, we are the first to test GloVe for system call representation and also to use a combination of system calls and kernel modules as input options.

\textbf{Structure.} 
The rest of this paper is organized as follows. 
Section \ref{sec:related} discusses related approaches for intrusion detection using system calls. 
Section \ref{sec:method} provides details on the four preprocessing approaches analyzed in this work.
Section \ref{sec:data} presents available datasets and their properties.
Experiments and results are presented in Sections \ref{sec:experiments}.
Finally, Section \ref{sec:conclusion} summarizes this work.   

\section{Related Work}
\label{sec:related}
This Section focuses on how sequences of system calls can be interpreted for intrusion detection and how the application of neural networks puts even more emphasis on the proper preprocessing of input data.

Starting with the work of Forrest et al. \cite{forrest1996sense}, the use of system call sequences for anomaly or intrusion detection has gained considerable interest.
Forrest et al. interpret system calls as categorical values and compare them with respect to equality or inequality.
They use a sliding window across system call traces of benign programs to identify the relative positions of system calls to each other and store them in a database.
%In each sliding window, the relative positions of the system calls to each other are stored in a database.
If a program is run, its trace is compared to the known relative positions from the database.
Accumulated deviations between the current trace and database data which exceed a threshold may indicate an attack.

Ever since a lot of research has been based on the use of system call sequences for intrusion detection. 
During the last decade, methods shifted from Hidden Markov Models \cite{eskin2001modeling,hoang2004efficient,kosoresow1997intrusion} and Support Vector Machines \cite{eskin2002geometric,wang2004anomaly} towards various kinds of neural networks \cite{chawla2018host,creech2014semantic,kim2016lstm}.
As neural networks can only process numerical data, a lot of effort was put into the meaningful transformation of system call sequences into numerical representations.
Popular methods from NLP were applied to represent the traces, since temporal ordering of input data is also very important for analyzing language.

Creech et al. \cite{creech2014semantic} create semantic models based on context-free grammars. 
Similar to Forrest \cite{forrest1996sense}, they create a database of normal behavior, but then use counting to generate input features for a decision engine.
The database is built by forming a dictionary containing every contiguous trace by using multiple sliding window sizes, denoting the resulting traces as words.
Those words (dictionary entries) are again combined to form phrases of length one to five. 
Finally, the counts of actual occurrences of those phrases are fed into different decision engines, e.g. a neural network called Extreme Learning Machine (ELM).
Xie et al. \cite{xie2014evaluating}, however, found that learning the dictionary is extremely time consuming, in particular an entire week for the ADFA-LD dataset. 

Murtaza et al. \cite{murtaza2015trace} aim at reducing the execution time of anomaly detectors by representing system call traces as sequences of kernel module interactions.
They map each single system call to its corresponding kernel module, thus reducing the range of input values from approximately 300 system calls to eight kernel modules.
This approach achieves similar or fewer false alarms and is much faster. 
Yet, the original seven kernel modules of a 32 bit Unix system (architecture, file systems, inter process communication, kernel,
memory management, networking, and security) need to be extended by another kernel module (unknown) to capture all possible system calls, since the system call table the authors use does not comprise every system call within the ADFA-LD.

While neural networks are widely used as classifiers, they can also be employed to extract a meaningful representation of the input data.
Two of the most popular approaches up to date are word2vec \cite{mikolov2013distributed} and GloVe \cite{pennington2014glove}.
Both methods are heavily used in NLP as they can extract a vector representation of words regarding their context of use.
%Both methods are explained in more detail in Section \ref{sec:method}.
Variants of word2vec have already been used by Ring et al. \cite{ring2017ip2vec} for network-based intrusion detection.

Following a similar idea, Kim et al. \cite{kim2016lstm} combine an additional embedding layer with a LSTM-based sequence predictor.
The authors state that through using an additional fully-connected layer in front of a LSTM layer, the sequence fed into the neural network becomes embedded to continuous space in a first step.
Since the authors use this LSTM-based approach for sequence prediction on the ADFA-LD dataset as well, we adopt the idea of adding an additional embedding layer as one potential (built-in) preprocessing approach.
The single system calls become embedded before reaching the LSTM layer.

\section{Comparison of System Call Representations}
\label{sec:method}
The overall purpose of this work is to compare four different methods of system call representation such that neural networks can process them.
Since isolated system calls do not contain any clue regarding their intent (benign or malicious), it is necessary to take their relationships into account by analyzing sequences of system calls.
%An isolated system call does not contain any clue regarding its intent (benign or malicious).
%Thus, it is necessary to analyze sequences of system calls to take their relationship into account.

\begin{figure}[htbp]
\centering%
\fbox{\includegraphics[scale=0.08]{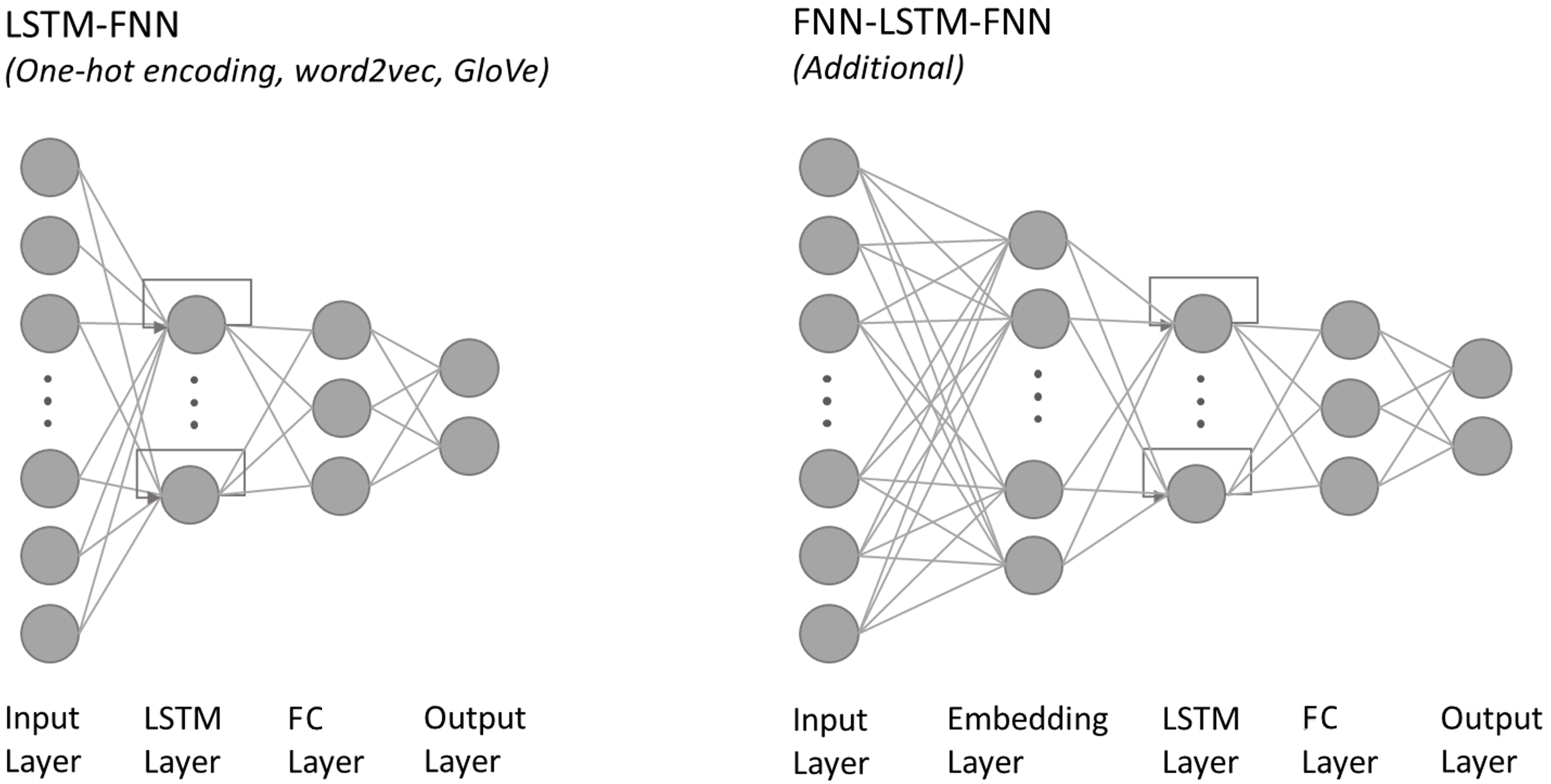}}
\caption{Network structures used.}\label{fig:lstm}
\end{figure}

Consequently, a simple LSTM classifier is used for comparing the four input methods.
The LSTM classifier used, in its original definition, consists of an input layer, a LSTM layer, a fully connected (FC) layer and an output layer.
Figure \ref{fig:lstm} on the left shows the used network structure for three of the four methods being compared.
The right side shows the network structure for the fourth method.

The main contribution is the comparison of system call representation techniques, in particular widely used methods from NLP.
The use of a LSTM for classification is a mere tool to show the effect different input methods have on the intent of a sequence (benign or malicious).
Again, our approach should not be seen as a solution to intrusion detection, but rather as a systematic comparison of different input possibilities.
%Since a simple sequence classifier is used, the effect of different input methods for differentiating sequences of system calls with respect to their nature (benign or malicious) should become more apparent.

Like Murtaza et al. \cite{murtaza2015trace}, we map system calls to their corresponding kernel modules.
%In contrast to \cite{murtaza2015trace}, we can do without the unknown kernel module by further analysis of the ADFA-LD dataset.
Since ADFA-LD was recorded on an Ubuntu 11.04 32 bit operating system, all system calls can be assigned to their respective kernel modules by looking into the system call implementation in the source files of kernel 2.6.35.
Thus, in contrast to \cite{murtaza2015trace}, no unknown kernel module is needed.

%Aside from the original system call sequences, we follow \cite{murtaza2015trace} in taking the kernel modules of a Unix system into account, but also use a combination of system calls and their corresponding kernel modules.
With this, three forms of input to a neural network emerge for a comprehensive comparison:
Using (1) only the system call representation,
(2) only the kernel module representation or
(3) both combined.
Consequently, this paper compares twelve different input variants of four categories as discussed below.

\textbf{One-hot encoding.} 
One-hot encoding is the typical approach for feeding categorical data into a neural network.
Each system call is represented with a vector in which every position represents a specific system call.
Hence, the size of the input vector equals the number of different system calls.
%In particular, in an one-hot encoding of system calls every position of the vector represents a specific system call.
A particular system call is mapped to a binary vector of all zeros except for a single one at the appropriate position for the corresponding call.
This method of representation will be tested using three forms:
(1) The system calls in one-hot encoding. 
(2) The mapped kernel modules in one-hot encoding.
(3) Both one-hot vectors (the system call vector of size 341 and the module vector of size 7) concatenated.
Any of these input techniques will be referred as \textit{One-hot} in the following.

\textbf{Additional Embedding Layer.}
This category accords with \textit{one-hot encoding}, except for an additional (fully-connected) embedding layer between input and LSTM layers (see right part of Figure 2).
Thus, the input will be embedded to a continuous vector representation using the embedding layer in front of the LSTM layer.
This approach is inspired by a similar network structure for sequence prediction \cite{kim2016lstm}.
As before, the following three forms of input will be used:
(1) The system calls in one-hot encoding. 
(2) The mapped kernel modules in one-hot encoding.
(3) Both one-hot vectors (the system call vector of size 341 and the module vector of size 7) concatenated.
These input techniques using the additional embedding layer, will be referred to as \textit{Additional} in this work.

\textbf{Word2vec.}
One-hot encoding has still many meaningful applications and is very popular due to its simplicity, but cannot cope with more complex structures like natural language text since the meaning of words may depend on their context.
Mikolov et al. \cite{mikolov2013distributed} presented an approach now known as word2vec which generates word vectors based on the context in which they are used.
Word vectors may be used following two approaches, namely the Continuous Bag-of-Words model (CBOW) and the Skip-Gram model.
CBOW learns to predict target words from given context (e.g. the context being 'Molly is already at' and the target word being 'home').
The Skip-Gram model works the other way around, predicting context from a target word.
The basic idea of word2vec is to train a neural network, discard the model, but use the weights of the fully trained hidden layer as word vectors.

Since system calls may also vary in their intent (benign or malicious) given their context around them, we adapt this NLP approach using the CBOW model.
Consequently, this category consists of the following three input methods:
(1) The system calls in word2vec representation. 
(2) The mapped kernel modules in word2vec representation. 
(3) The concatenation of system calls in word2vec representation and one-hot encoded kernel modules.
This input category will be referred to as \textit{word2vec} in this work.

\textbf{GloVe.}
GloVe \cite{pennington2014glove} is a count-based model that also learns a vector representation of words regarding their context.
For GloVe, context amounts to a co-occurrence matrix, thus including word statistics into their model.
GloVe is then trained with the non-zero entries of that co-occurrence matrix.
As with word2vec, this category consists of the three input methods:
(1) The system calls in GloVe representation. 
(2) The mapped kernel modules in GloVe representation. 
(3) The concatenation of system calls in GloVe representation and one-hot encoded kernel modules.
These three forms of input will be referred to as \textit{GloVe} in the remaining chapters.
\section{Data}
\label{sec:data}
Three well-known datasets, namely the DARPA \cite{DARPA}, UNM \cite{UNM} and ADFA-LD \cite{creech2013generation}, are very popular for evaluating and comparing host-based intrusion detection systems. 
A typical system call sent to the kernel consists of its name, parameters and a return value.
Since number and type of the transfer values vary among system calls, many researchers focus on analyzing only the temporal ordering of system calls, i.e. their names, but neglecting parameters to reduce complexity \cite{creech2014semantic,forrest1996sense,warrender1999detecting,xie2014evaluating}.
Hence, popular datasets like the UNM and ADFA-LD only take the system calls names into account.
 
%To this day the DARPA and UNM datasets are over 15 years old and are still being used due to a lack of better alternatives.
The DARPA and UNM datasets are already more than 15 years old, but still in use due to a lack of better alternatives.
However, they are also heavily criticised due to their age and lack of complexity \cite{mchugh2000testing}.
Creech et al. \cite{creech2013generation} recorded the ADFA-LD dataset in order to replace the outdated DARPA (KDD) collection.
Hence we focus solely on the newer ADFA-LD set recorded in 2013.

The ADFA-LD dataset consists of three parts named \textit{Training\_Data\_Master}, \textit{Attack\_Data\_Master} and \textit{Validation\_Data\_Master}, which contain files with system call traces of processes.
Table \ref{adfaset} shows the number of system calls and traces contained in the three dataset partitions together with the type of behavior.

\setlength{\tabcolsep}{4pt}
\renewcommand{\arraystretch}{1.0}
\captionsetup{font={footnotesize,sc},justification=centering,labelsep=period}%
\begin{table}[htbp]
\caption{Number of System Calls and Traces in ADFA-LD.}\label{adfaset}
\centering%a
\begin{tabular}{llll}
\hline
\textit{} & \textit{System Calls} & \textit{Traces} & \textit{Label of Traces} \\
\hline
\textit{Training} & 308077 & 833 & benign\\
\textit{Attack} & 317388 & 746 & malicious \\
\textit{Validation} & 2122085 & 4372 & benign\\
\hline
\end{tabular}
\end{table}
\captionsetup{font={footnotesize,rm},justification=centering,labelsep=period}%

Only the \textit{Attack\_Data\_Master} contains exploited processes. 
In order to use the ADFA-LD dataset for classification, the attack subset needs to be split, mixing one half with the training and the other half with the validation subset.
Since training and attack subsets encompass roughly the same number of system calls, the new combined training dataset is skewed with respect to its class label after splitting the attack subset.
The new set contains around twice as many benign sequences as malicious sequences.
This is even worse in the validation data set since splitting results in a combination of the original roughly 2000000 system calls from benign traces and approximately 150000 system calls from malicious traces. 
If such a skewed model classifies every input as benign, an accuracy of 93\% will be achieved even without any attacks detected.
Working with these unbalanced sets demands two adaptations: 
(1) To avoid a resulting model which tends to classify sequences as benign, the new training set is balanced by using data point duplication in the less represented class. This technique is called random oversampling \cite{he2008learning}.
(2) Since the new combined validation dataset is extremely unbalanced, accuracy is no clear indicator towards the quality of the sequence classifier.
Therefore, true positive rates and false positive rates (TPR and FPR) are used as evaluation criteria.

Training the LSTM model using a whole process trace file does not make much sense due to the nature of attacks.
An intrusion detection system should not wait until an exploit reaches its end, but rather intervene early, e.g. by stopping the exploited process.
A typical approach (see Section \ref{sec:related}) is to use sliding windows to split the dataset into smaller sequences for training, attaching the corresponding label of the trace.
A process trace is divided into smaller sequences that can be processed without waiting for the process to end.
In this work, a sliding window size of 20 is used due to the structure of the ADFA-LD.
%While the \textit{Attack\_Data\_Master} consists of traces of exploited processes, that does not necessarily mean that each cut out sequence within is showing malicious behaviour, but it still would be labeled by the intent of the process (malicious).
The \textit{Attack\_Data\_Master} consists of traces of exploited processes. 
Yet, not every partial sequence exhibits malicious behavior, but it still would be labeled by the overall intent of the process (malicious). 
For example, a process may start normally and an attack could exploit a vulnerability near the end of the trace. 
So, to reduce the risk of learning malicious labels for benign sequences we refrain from using typical smaller window sizes like 5, 6 or 11 as used in \cite{forrest1996sense}.
\section{Experiments}
\label{sec:experiments}

\subsection{Experiment Setup}
\label{sec:experimentsetup}
Table \ref{parameter} shows the parameters used for the neural networks.
For a fair comparison, the same overall parameters are used in all settings. 
Embedding sizes are generally set to the size of 8 except when using kernel modules only.
%Since there are only seven different kernel modules present, using the same embedding size, LSTM size and fully-connected layer (FC) size is not reasonable (in comparison to a size of 341 different system calls or 348 system calls and kernel modules combined).
Since there are only seven different kernel modules compared to 341 different system calls or 348 system calls and kernel modules combined, using identical sizes for embedding, LSTM and fully-connected (FC) layer is not reasonable.
Thus, an embedding size and FC size of 3 is used when working with kernel modules only.
In Table \ref{parameter}, text embedding parameters like the vector size of word2vec or GloVe are indicated as TE (text embedding). 
Text embeddings are learned for 10 epochs.

\setlength{\tabcolsep}{4pt}
\renewcommand{\arraystretch}{1.1}
\captionsetup{font={footnotesize,sc},justification=centering,labelsep=period}%
\begin{table}[htbp]
\caption{Parameterset (N.A. -- not applicable).}\label{parameter}
\centering%
\begin{tabular}{p{1.4cm}p{1.7cm}p{1.2cm}p{1.2cm}p{1.2cm}p{1.2cm}p{1.2cm}}
\hline
\textit{Source} & \textit{Method} & \textit{TE Size} & \textit{TE Window} & \textit{Embed Size} & \textit{LSTM Size} & \textit{FC Size} \\
\hline
\multirow{3}{*}{\shortstack[l]{System\\ Calls}} 
 & One-Hot & N.A. & N.A. & N.A. & 32 & 16 \\
 & Additional & N.A. & N.A. & 8 & 32 & 16\\
 & Word2vec & 8 & 5 & N.A. & 32 & 16 \\
 & GloVe & 8 & 5 & N.A. & 32 & 16 \\ \hline
\multirow{3}{*}{\shortstack[l]{Kernel\\ Modules}} 
 & One-Hot & N.A. & N.A. & N.A. & 5 & 3 \\
 & Additional & N.A. & N.A. & 3 & 5 & 3 \\
 & Word2vec & 3 & 5 & N.A. & 5 & 3 \\
 & GloVe & 3 & 5 & N.A. & 5 & 3 \\ \hline
\multirow{3}{*}{Both} 
 & One-Hot & N.A. & N.A. & N.A. & 32 & 16 \\
 & Additional & N.A. & N.A. & 8 & 32 & 16 \\
 & Word2vec & 8 & 5 & N.A. & 32 & 16 \\
 & GloVe & 8 & 5 & N.A. & 32 & 16 \\
\hline
\end{tabular}
\end{table}
\captionsetup{font={footnotesize,rm},justification=centering,labelsep=period}%

Dropout is a general strategy to avoid overfitting in neural networks \cite{srivastava2014dropout}, whereas peepholes are an optimization specifically for the timing in LSTM cells \cite{gers2002learning}.
Here, dropout is set using a keeping probability of 0.8 for all experiments on the final FC layer, and peepholes in the LSTM layers are set to true.

As mentioned in Section \ref{sec:data}, we use a sequence length of 20.
For fair comparison, each model is trained the same amount of epochs (20).
Naturally, the number of classes (benign, malicious) implies a size of 2 for the output layer.

\subsection{Evaluation of Results}
\label{sec:results}
Table \ref{restab} shows the experiment results.
Since the dataset is skewed in its class distribution, higher accuracy does not necessarily mean a better result.
As explained above, true positive rates (TPR) and false positives rates (FPR) might be more meaningful.
In our setting, a true positive is a correctly detected malicious sequence while a false positive is a normal sequence that is classified to be malicious.
To reduce confounding effects, we conduct our experiments twice on the ADFA-LD dataset with different splits.
Table \ref{restab} shows the mean results of both experiments.
\setlength{\tabcolsep}{4pt}
\renewcommand{\arraystretch}{1.5}
\captionsetup{font={footnotesize,sc},justification=centering,labelsep=period}%
\begin{table}[htbp]
\caption{Results: True Positive Rate/False Positive Rate (Accuracy).}\label{restab}
\centering%
\begin{tabular}{p{1.5cm}p{1.5cm}p{1.5cm}p{1.5cm}p{1.5cm}}
\hline
\textit{} & \textit{One-hot} & \textit{Additional} & \textit{Word2vec} & \textit{GloVe} \\
\hline
\textit{System Calls} & 0.95/0.16 (0.85)
						& 0.90/0.16 (0.85)
                        & 0.92/0.17 (0.84)
                        & 0.79/0.14 (0.85) \\
\textit{Kernel Modules} & 0.80/0.24 (0.77)
						& 0.89/0.25 (0.75)
                        & 0.77/0.25 (0.76)
                        & 0.77/0.24 (0.77) \\
\textit{Both}        &  0.95/0.16 (0.86)
						& 0.93/0.17 (0.84)
                        & 0.91/0.16 (0.85)
                        & 0.87/0.16 (0.84) \\
\hline
\end{tabular}
\end{table}
\captionsetup{font={footnotesize,rm},justification=centering,labelsep=period}%

There is always the risk that valuable information regarding the intent of a sequence (benign or malicious) is lost during preprocessing steps.
In our case, this does not seems to be the case for all four transformation methods (one-hot, additional, word2vec or GloVe).
Nevertheless, one-hot encoding, which is the most direct transformation of the non-continuous system calls, achieves the best results with a TPR of 0.95 and a FPR of 0.16.

Our results are in line with other intrusion detection based results on this dataset.
%To put the results presented in Table \ref{restab} into perspective, two approaches of authors who were partly involved in the recording of the ADFA-LD dataset should be mentioned.
The resulst of Creech et al. \cite{creech2014semantic} show a TPR of approximately 90\% at a FPR of around 15\%.
Xie et al. \cite{xie2014evaluating} achieve a TPR of 70\% at a FPR of around 20\%.
%It should be noted, though, that those two approaches are anomaly-based, hence they cannot be directly compared to our classification-based results, but should merely give an indication about the range our results can be compared to.
It should however be noted, that those two approaches are anomaly-based (hence they can not be directly compared to our classification-based results), but even if the setting is not comparable, the similar range shows that the representations work quite well.
%Again, our approach should not be seen as a solution to intrusion detection, but rather as a systematic comparison of different input possibilities.

However, using kernel modules only results in a considerably lower TPR and higher FPR in comparison to our other approaches and is, thus, not advised to use in this classification setting.
Surprisingly, enriching the system call data with their corresponding kernel modules does not yield better results.
Also, neither the additional embedding layer nor the pre-learned NLP representations show better results than one-hot encoding without the additional embedding layer.
This might be because of the nature of LSTM cells.
Since LSTM cells automatically embed sequences of data, additional embedding prior to the LSTM might not be helpful.
However, aside from using kernel modules only, the overall information regarding the intent of the sequences is being kept to a certain extent.
So far, we used the same overall parameters for all approaches in order to have a fair comparison base.
Better results could be achieved for each approach with parameter optimization.
By and large, each of the four methods seem to be rewarding in their setting.

Which method from Table \ref{restab} is considered to be the best also heavily relies on the problem setting at hand.
In a setting of intrusion detection through classification, we would argue that a small false positive rate is more important than a high true positive rate for two reasons.
(1) Analyzing an alert with respect to it being a true or false positive is very expensive.
(2) It may not even be necessary to get all malicious sequences, since it is sufficient to raise an alert on one of the malicious system call sequences within one exploited process trace.
The latter again depends on the problem setting or rather how to handle an attack reported. 
For instance, in critical infrastructures it may be more important to capture every attack possible.
In this case, analyzing potential false positives may be a necessary evil.
\section{Conclusion}
\label{sec:conclusion}
This work systematically compares different input methods for system call traces for intrusion detection.
%Three different options for representing system calls where combined with those input methods.
We use sequences of system calls, their mapped kernel modules or a combination of both as representation options.
The three input options are combined with four different encodings, namely an one-hot vector, learning an embedding while training, and using word2vec or GloVe representations.

Results imply that working with kernel modules exclusively is not recommended for our setting, although they might still be helpful as supplementary information in other settings.
One-hot encoding showed the best results.
The other approaches, however, should not be discarded since they could be helpful for more sophisticated models for intrusion detection.
Also we compare every approach with the same overall parameterset to achieve a fair comparison base. 
Better results in terms of TPR or FPR might be achieved by optimizing a specific approach.
%Then, for example, when modeling normal behavior for anomaly detection, adding context information using word2vec representation might still be helpful.
%Important is that the intent of a sequence is not becoming ambiguous during a preprocessing step.
%So, this work analyzes what impact different preprocessing methods have on sequences of system calls regarding their intent. 
%This is achieved by comparing the different methods using a classification task for benign or malicious behavior on the ADFA-LD dataset.
%However, as can be seen in the different approaches summarized in Section \ref{sec:related}, anomaly detection is preferred to normal classification tasks in intrusion detection.
%Still, preprocessing is an important step when using neural networks for an anomaly detection approach.
%It is important, that the intent of a sequence to be modeled as normal behavior remains clear after applying preprocessing steps. 
%So, with the results of this work in mind, future activities may focus on modeling normal behavior of programs based on system call sequences.

With the results of this work in mind, future activities may focus on modeling normal behavior of programs based on system call sequences.

\textbf{Acknowledgements.}
S.W. is funded by the Bavarian State Ministry of Science and Arts in the framework of the Centre Digitization.Bavaria (ZD.B).
S.W. and M.R. are further supported by the BayWISS Consortium Digitization.
Last but not least, we gratefully acknowledge the support of NVIDIA Corporation with the donation of the Titan Xp GPU used for this research.\newline\newline
%\textit{This is a pre-print of a contribution sent to the 12th International Conference on Computational Intelligence in Security for Information Systems (CISIS 2019).}
This is a pre-print of a contribution published in Advances in Intelligent Systems and Computing (AISC, volume 951) published by Francisco Martínez Álvarez, Alicia Troncoso Lora, José António Sáez Muñoz, Héctor Quintián and Emilio Corchado.
The definitive authenticated version is available online via https://doi.org/10.1007/978-3-030-20005-3\_2.

\bibliographystyle{spmpsci}
\bibliography{literature}
\end{document}